\documentclass{aa}  
\usepackage{graphicx}
\usepackage{ulem}

\usepackage{txfonts}

\usepackage{hyperref}
\hypersetup{
	colorlinks=true,
	linkcolor=blue, 
	urlcolor=blue,
	citecolor=blue,
}
\usepackage{multirow}
\usepackage{CJK}

\titlerunning{Selected open cluster sample for validating atmospheric parameters}
\authorrunning{Tang et al}

\begin{document} 
\begin{CJK*}{UTF8}{gbsn}

   \title{Selected open cluster sample for validating atmospheric parameters: Application to Gaia and other surveys}

\author{Tong Tang(唐通)
\inst{1,2}
          \and
          Songmei Qin(秦松梅)
\inst{1,2,3}
          \and
         Jing Zhong(钟靖)
\inst{1}
          \and
          Yueyue Jiang(蒋悦悦)
\inst{1,2}
          \and
          Li Chen(陈力)
\inst{1,2}}

\institute{ Astrophysics Division, Shanghai Astronomical Observatory, Chinese Academy of Sciences, 80 Nandan Road, Shanghai 200030, PR China
        \and
        School of Astronomy and Space Science, University of Chinese Academy of Sciences, No. 19A, Yuquan Road, Beijing 100049, PR China
        \and
        Institut de Ci\`encies del Cosmos, Universitat de Barcelona (ICCUB), Mart\'i i Franqu\`es 1, 08028 Barcelona, Spain\\
        \email{jzhong@shao.ac.cn} }

  \abstract
   {Reliable stellar atmospheric parameters are essential for probing stellar structure and evolution, and for stellar population studies. However, various deviations appear in comparisons with different ground-based spectroscopic surveys.}
   {We aim to select high-quality open cluster members and employ the atmospheric parameters provided by the theoretical isochrones of open clusters as a benchmark to assess the quality of stellar atmospheric parameters from {\it Gaia} DR3 and other ground-based spectroscopic surveys, such as LAMOST DR11, APOGEE DR17, and GALAH DR4.}
   {We selected 130 open clusters with well-defined main sequences within $500\,\rm pc$ of the solar neighborhood as a benchmark sample to estimate the reference atmospheric parameters of the members from the best-fit isochrones of those clusters.}
   {By comparing the atmospheric parameters provided by different spectroscopic surveys to the theoretical parameters, we found that the atmospheric parameter deviation and the corresponding dispersions exhibit different variations. The atmospheric parameter deviations of F, G, and K-type stars are smaller than those of B, A, and M-type stars for most surveys. For most samples, the dispersion of $T_{\rm eff}$ decreases as temperature decreases, whereas the dispersion of $\log g$ shows the opposite trend.}
   {}

  \maketitle

\defcitealias{2023A&A...674A..28F}{F23}
\defcitealias{2023A&A...677A.162B}{B23}
\defcitealias{2023ApJS..265...12Q}{Qin23}

\section{Introduction}

Stellar spectra allow us to obtain basic stellar atmospheric parameters, such as the effective temperature ($T_{\rm eff}$), the surface gravity (log $g$), and the metallicity ([M/H]) \citep{2011RAA....11..924W,2014A&A...569A.111B}. This information is essential not only for exploring stellar formation and evolution but also for understanding the structure and formation history of the Milky Way \citep{2022A&A...668A...4F,2022MNRAS.509..421N}.

The ambitious European Space Agency (ESA) mission {\it Gaia} \citep{2016A&A...595A...1G} has provided insights into the physical properties of Milky Way stars. {\it Gaia} Data Release 3 (DR3, \citealt{2023A&A...674A...1G}) published high-precision astrometric, photometric, and atmospheric parameters for a vast number of stars, which revolutionized the study of stars and the Milky Way. The General Stellar Parameterizer from Photometry (GSP-Phot) is an important module that aims to estimate atmospheric parameters from low-resolution blue photometer (BP) and red photometer (RP) spectra. It provided the atmospheric parameters, including $T_{\rm eff}$, $\log{g}$, and [M/H], for about 470 million sources \citep{2023A&A...674A..26C,2023A&A...674A..27A}. Meanwhile, {\it Gaia} DR3 has released low-resolution BP/RP spectra ($\lambda / \Delta \lambda = 20-60$) covering wavelength ranges of $330-680$\,\rm nm and $640-1050$\,\rm nm for approximately 220 million stars with sufficient observation times and high signal-to-noise ratios \citep{2023A&A...674A...2D}. The General Stellar Parameterizer from Spectroscopy (GSP-Spec) is another crucial module of the {\it Gaia} mission, estimating the chemophysical parameters for about 5.6 million stars using the pure spectroscopic processing of the Radial Velocity Spectrometer (RVS) ($\lambda / \Delta \lambda \sim 11\,\rm 500$), which covers the wavelength range of $846-870$\,\rm nm \citep{2023A&A...674A..29R}. Moreover, the Extended Stellar Parameterizer for Hot Stars (ESP-HS), a module dedicated to analyzing high-temperature stars (O, B, and A-type stars), provided atmospheric parameters for about 2.4 million stars \citep[hereafter \citepalias{2023A&A...674A..28F}]{2023A&A...674A..28F}. It is noted that the Gaia methods do not incorporate prior information like binary stars, member stars, or known distances when estimating these atmospheric parameters on a star-by-star basis from the observational data. Also, in dense regions there could be issues in the BP, RP, and RVS  astrometric data because of the limits in the angular resolution.

Other spectroscopic surveys also provide a large sample of atmospheric parameters, such as the Large Sky Area Multi-Object Fiber Spectroscopic Telescope survey (LAMOST, \citealt{2012RAA....12.1197C,2012RAA....12..723Z}), the Apache Point Observatory Galactic Evolution Experiment survey (APOGEE, \citealt{2017AJ....154...94M}), and the GALactic Archaeology with HERMES survey (GALAH, \citealt{2021MNRAS.506..150B}). 
The LAMOST Low-Resolution Spectroscopic Survey provided a stellar parameter catalog for about 7.7 million A, F, G, and K-type stars (LAMOST-LRS DR11 v1.0\footnote{\url{https://www.lamost.org/dr11/v1.0}}) from the LAMOST Stellar Parameter Pipeline (LASP, \citealt{2011RAA....11..924W}). The published low-resolution spectra ($\lambda / \Delta \lambda \sim 1800$) cover the wavelength range of $370-900$\,\rm nm.
The most current APOGEE data release 17 (DR17, \citealt{2022ApJS..259...35A}) includes spectroscopic parameters determined by the APOGEE Stellar Parameters and Chemical Abundances Pipeline (ASPCAP, \citealt{2016AJ....151..144G}) for more than 730\,\rm 000 stars with high-resolution spectra ($\lambda / \Delta \lambda \sim 22\,\rm 500$, $1510-1700$\,\rm nm).
The fourth data release (DR4, \citealt{2024arXiv240919858B}) of the GALAH survey contains stellar parameters provided by Spectroscopy Made Easy (SME, \citealt{2012ascl.soft02013V}) for 917\,\rm 588 stars from high-resolution spectra with four noncontiguous wavelength bands in the range of $471-490$, $565-587$, $648-674$, and $759-789$\,\rm nm.

Several recent studies have compared the {\it Gaia} atmospheric data with other survey data. \citet{2023A&A...674A..27A} compared GSP-Phot parameters to those from APOGEE DR16, GALAH DR3, LAMOST DR4, and RAVE DR6, obtaining median absolute differences of 169\,\rm K, 110\,\rm K, 110\,\rm K, and 160\,\rm K for $T_{\rm eff}$ and 0.22\,\rm dex, 0.06\,\rm dex, 0.1\,\rm dex, and 0.25\,\rm dex for $\log{g}$, respectively. They suggested that GSP-Phot tends to overestimate $T_{\rm eff}$ in the Galactic plane (see their Fig. 8) and systematically overestimates log $g$ (see their Fig. 9).
\citet{2023A&A...674A..29R} compared GSP-Spec parameters with APOGEE DR17, GALAH DR3, and RAVE DR6 with a selected best-quality sample. They determined a median offset for $T_{\rm eff}$ and $\log{g}$ of $-$17\,\rm K and $-$0.3\,\rm dex, respectively.
\citetalias{2023A&A...674A..28F} compared the ESP-HS parameters for hot stars with some literature catalogs. They found that the dispersions in $T_{\rm eff}$ and $\log{g}$ deviation values between the ESP-HS and other catalogs increased with temperature. These parameters obtained from different spectroscopic surveys vary in the observed band, spectral resolution, and data processing method, which may lead to some systematic differences. For example, the deviation and dispersion of $T_{\rm eff}$ between GSP-Phot and APOGEE DR16 are $2-3$ times greater than those between GSP-Phot and LAMOST DR4, GALAH DR3, and RAVE DR6. Therefore, it is a challenge to accurately assess the quality of the atmospheric parameters of the different surveys. 

To test the diverse systematic deviations among different spectroscopic surveys, we employed the atmospheric parameters from PARSEC isochrones \citep{2017ApJ...835...77M} as a benchmark to assess the observational atmospheric parameters of {\it Gaia} DR3, LAMOST-LRS DR11, APOGEE DR17, and GALAH DR4. Compared with field stars, it is noted that open clusters can be used as a bridge between observation and theoretical parameters, which are highly efficient for evaluating the quality of stellar parameters \citep{2020A&A...640A.127Z,2022A&A...668A...4F,2023A&A...674A..28F}. This is because the member stars in an open cluster are formed in the same molecular cloud, sharing a similar age, metallicity, distance, and extinction \citep{2003ARA&A..41...57L,2010ARA&A..48..431P}, and so can be obtained with a more accurate age through isochrone fitting \citep{2010A&A...516A...2M,2019A&A...623A.108B}. After determining the stellar age, their parameter values, such as the $T_{\rm eff}$ and $\log{g}$ of cluster members, can be estimated by comparing the PARSEC isochrones and their locus on the color-magnitude diagrams (CMDs).

The unprecedented precise astrometric and photometric from {\it Gaia} data \citep{2023A&A...674A...1G}, together with the popularity of machine learning approaches \citep{2014A&A...561A..57K,2021A&A...650A.109P,2017JOSS....2..205M,2021A&A...646A.104H} in finding the open cluster, has effectively boosted the cluster census and the reliability of membership determination. More than 4000 open clusters have been detected in the Milky Way \citep{2018A&A...618A..93C,2020A&A...640A...1C,2019ApJS..245...32L,2019JKAS...52..145S,2020A&A...635A..45C,2022A&A...661A.118C,2021RAA....21...45Q,2023ApJS..265...12Q,2022ApJS..260....8H,2022ApJS..262....7H,2023A&A...673A.114H}, which offers us an opportunity to evaluate the atmospheric parameters from {\it Gaia} DR3 with a large and reliable sample of member stars.

\citetalias{2023A&A...674A..28F} estimated reference values of atmospheric parameters from isochrone fitting for 230\,\rm000 cluster members and compared the atmospheric parameters provided by {\it Gaia} DR3. They obtained a median of residuals to the isochrones for $T_{\rm eff}$, $\log{g}$ from GSP-Phot of $34\,\rm K$, $0.01\,\rm dex$, respectively, and a mean absolute deviation (MAD) of $400\,\rm K$, $0.22\,\rm dex$. They also pointed out the overestimation of $T_{\rm eff}$ for giants and $\log{g}$ for the main sequence stars and the underestimation of $T_{\rm eff}$ for supergiants and $\log{g}$ for hot stars and giants. When comparing the atmospheric parameters of GSP-Spec with the reference values, they found a median and a MAD of $T_{\rm eff}$, $\log{g}$ residuals to the isochrones of $6\,\rm K$ and $160\,\rm K$, $0.3\,\rm dex$ and $0.44\,\rm dex$, respectively.  
They also noted a systematic underestimation of both $T_{\rm eff}$ and $\log{g}$ provided by ESP-HS. Upon reviewing their sample of open clusters, we found that about 20\% exhibit apparent broadened main sequence features; for example, NGC 6124 and NGC 7654. This broadening could be caused by photometric uncertainties, differential reddening, binary stars, and so on, which may lead to biases in estimating the reference atmospheric parameters of the member stars. Hence, we need to select high-quality clusters with well-defined main sequences and exclude their binary member stars to reevaluate the quality of the atmospheric parameters provided by {\it Gaia}.

\citet[hereafter \citetalias{2023A&A...677A.162B}]{2023A&A...677A.162B} benchmarked the atmospheric parameters of {\it Gaia} DR3 using single stars in two well-known open clusters, Hyades and Pleiades. They indicated that the $T_{\rm eff}$ and $\log g$ of GSP-Phot deviate from the isochrone model by $200\,\rm K$ and 0.1-0.2$\,\rm dex$, with a dispersion of around $250\,\rm K$ and $0.2\,\rm dex$, respectively. The $T_{\rm eff}$ of the GSP-Spec is generally consistent with the isochrone, but the $\log g$ deviates significantly from the isochrone (Fig. 6 in \citetalias{2023A&A...677A.162B}). They also found an underestimation of the metallicity from GSP-Phot and GSP-Spec.

In our recent work, we systematically searched for open clusters in the Milky Way at Galactic latitudes of $|b|\leq30^\circ$ within $500\,\rm pc$ of the solar neighborhood using {\it Gaia} DR3 data \cite[hereafter \citetalias{2023ApJS..265...12Q}]{2023ApJS..265...12Q}. We employed varying slicing box sizes in different distance grids to distinguish the signals of cluster members from field stars. By combining the clustering algorithms pyUPMASK (Unsupervised Photometric Membership Assignment in Stellar Clusters, ~\citealt{2014A&A...561A..57K,2021A&A...650A.109P}) and HDBSCAN (Hierarchical Density-Based Spatial Clustering of Applications with Noise, ~\citealt{2017JOSS....2..205M}), a total of 324 open clusters were identified, including 101 previously unreported clusters. We also provided each open cluster's age, distance modulus, and reddening through visual isochrone fitting. 

In this work, we have selected 130 open clusters with clear and well-defined main sequences from the OCSN (Open Clusters of Solar Neighborhood) catalog provided by \citetalias{2023ApJS..265...12Q}. Considering the low extinction and high signal-to-noise ratio of these nearby open star clusters, which make it easier to measure more accurate atmospheric parameters, we use the member stars of these clusters to evaluate the quality of the stellar atmospheric parameter measurements from {\it Gaia} DR3, LAMOST-LRS DR11, APOGEE DR17, and GALAH DR4. 

The paper is structured as follows. In Sect.~\ref{sec:SM}, we describe the sample selection and the reference atmospheric parameter estimation process with PARSEC isochrones. In Sect.~\ref{sec:Vali-AP}, we provide the assessment of atmospheric parameters from {\it Gaia} and other spectroscopic survey. Finally, we sum up in Sect.~\ref{sec:C}.

\begin{figure*}
\includegraphics[width=0.98\textwidth, angle=0]{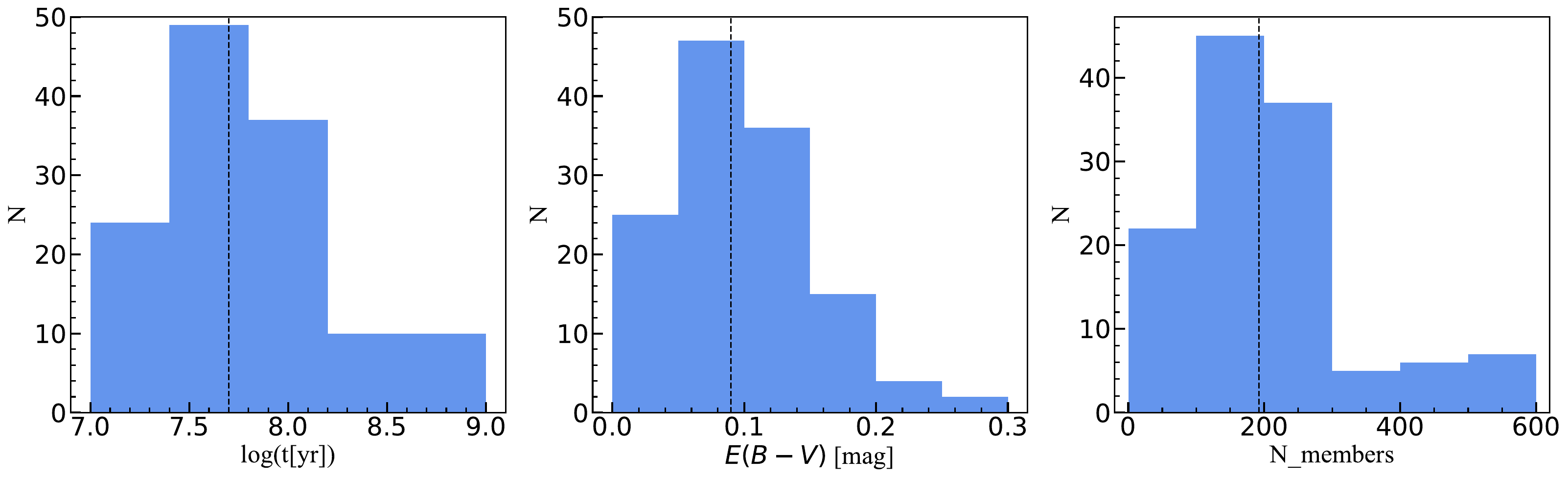}
\caption{Histograms of the selected clusters' age, reddening, and member number. Those parameters are from the OCSN catalog of \citetalias{2023ApJS..265...12Q}. The dashed black lines represent the median values.}
\label{Fig1}
\end{figure*}

\section{Sample and method}\label{sec:SM}
\subsection{Sample} \label{sec:Sample}

Open clusters are generally considered simple stellar populations, and all members in a cluster are supposed to present an isochrone distribution on the CMD. However, differential reddening and observational uncertainty would make the isochrone distribution become broader, leading to uncertainties when estimating the atmospheric parameters of members on the CMD. To reduce the effect of these factors, we selected open clusters with $E(B-V) \leq 0.3$ from the OCSN catalog of \citetalias{2023ApJS..265...12Q}, and then visually excluded clusters with an extended main sequence. A sample of 130 open clusters was selected, including 34 424 member stars in {\it Gaia} DR3. Fig.~\ref{Fig1} shows histograms of the age, reddening, and number of selected clusters provided by \citetalias{2023ApJS..265...12Q}. The logarithm ages of these clusters (log(t[yr])) are between 7 and 9; most are young. The number of members of each cluster ranges from 46 to 1986, with an average value of 269.

\begin{figure}[!htbp]
\includegraphics[width=0.49\textwidth]{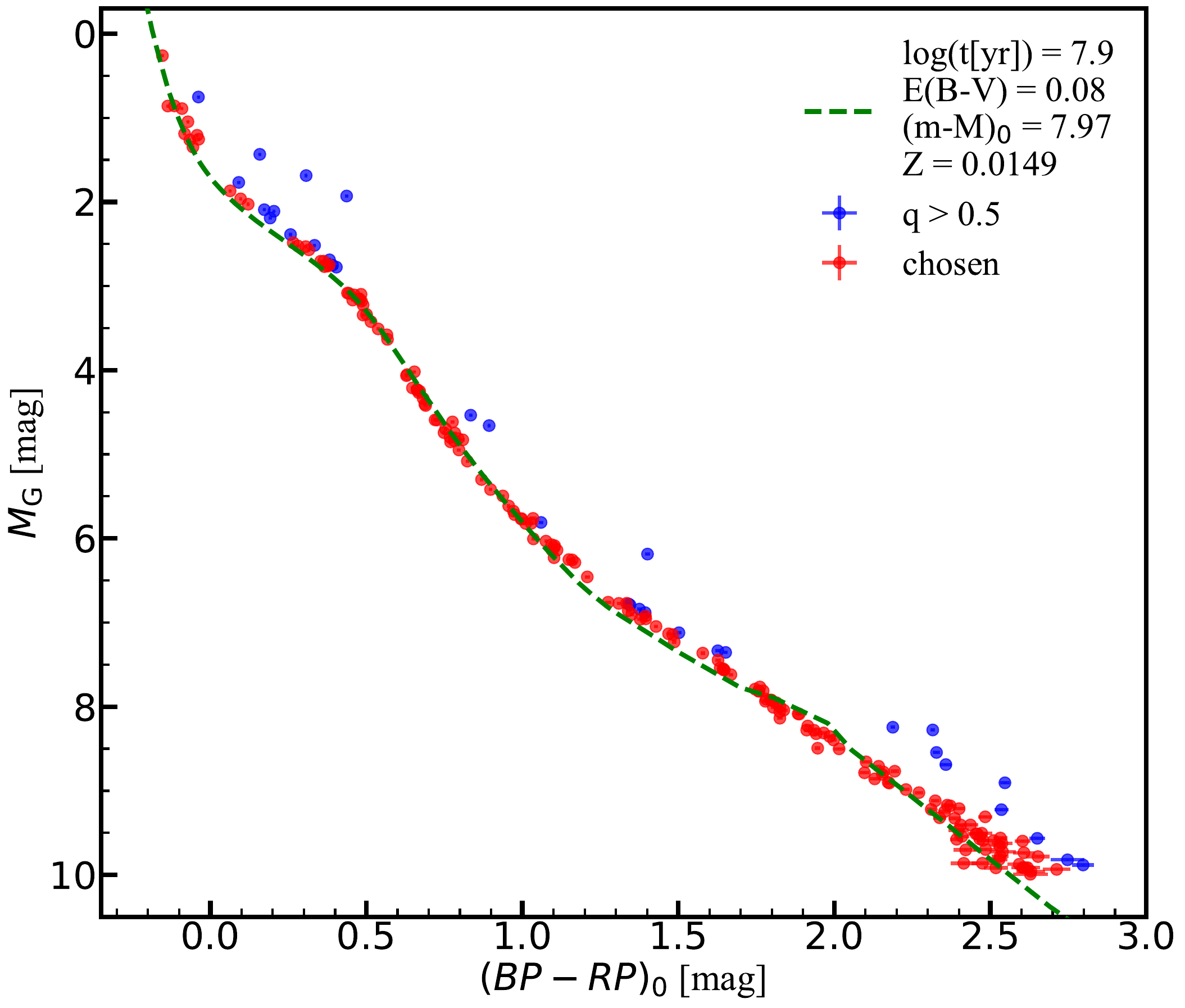}
\caption{Color-absolute magnitude diagram of OCSN 259 (Roslund 6). The dashed green line represents the best-fit isochrone provided by \citetalias{2023ApJS..265...12Q}. The blue dots represent the members with a binary mass ratio larger than 0.5. The red dots represent the member stars we have retained. The error bars indicate the photometric uncertainties.}
\label{Fig2}
\end{figure}

It is noted that \citetalias{2023ApJS..265...12Q} set the cut on the renormalized unit weight error as $<$ 1.4 \citep{LL:LL-124} to exclude sources with unreliable astrometric and photometric observations. Meanwhile, \cite{2021A&A...649A...3R} introduced the corrected flux excess factor of the BP and RP flux, $C^*$, defined as
\begin{equation}
C^* = C - f(BP-RP)
  \label{equ:C_func}
,\end{equation}
where $C=(I_{BP}+I_{RP})/I_G$ is the BP and RP flux excess factor, and $f(BP-RP)$ is a function indicating the expected excess at a given color for sources with high-quality photometry \citep{2021A&A...649A...3R}. 
We used $|C^*|<N \sigma_{C^*}$ to remove sources with inconsistent $G$-band photometry and BP and RP photometric measurements, where N=3, $\sigma_{C^*}$ in bins of $0.01\,\rm mag$ and fit with a simple power law in $G$ magnitude ( equation (18) in \citealt{2021A&A...649A...3R}). Furthermore, we made use of the best isochrones fit by \citetalias{2023ApJS..265...12Q} for all the clusters to derive their binary sequences with a mass ratio (q) of 0.5. Subsequently, we excluded those members that have a mass ratio greater than 0.5 (q > 0.5), as is represented by the blue dots in Fig.~\ref{Fig2}. 32\% of the member samples were removed in this step.

To validate the atmospheric parameters from spectroscopic surveys, we cross-matched the remaining cluster members with {\it Gaia} DR3 (GSP-Phot, GSP-Spec, ESP-HS), LAMOST-LRS DR11, APOGEE DR17, and GALAH DR4 atmospheric parameter catalogs. The atmospheric properties of the common samples are listed in Table~\ref{tab:sample}. Additionally, for the GSP-Spec sample, we applied the quality flags where vbroadT, vbroadG, VbroadM, vradT, vradG, and vradM were equal to 0; that is, f1, f2, f3, f4, f5, and f6=0 in the corresponding gspspec\_flags \citep{2023A&A...674A..29R}.

\begin{table}[!htbp]
\renewcommand{\arraystretch}{1.3}
\caption{Sample size and typical uncertainties from {\it Gaia}, LAMOST, APOGEE, and GALAH.}\label{tab:sample}
\centering
\Large 
\scalebox{0.6}{
\begin{tabular}{lccc}
\hline\hline
Catalog  & $N$ &  Uncertainty of $T_{\rm eff}$ &Uncertainty of $\log g$  \\ 
 &  & (K) & (dex)\\
\hline
GSP-Phot & 17\,\rm366 & 21 & 0.02  \\
GSP-Spec & 455 & 243 & 0.22  \\
ESP-HS &  1044& 71 & 0.03  \\
LAMOST-LRS DR11 &  777& 34 & 0.05  \\
APOGEE DR17 &  712& 12 & 0.02  \\
GALAH DR4 &  919& 69 & 0.10  \\
\hline
\end{tabular}}
\tablefoot{
For {\it Gaia} data, the typical uncertainty is defined as the median value of the upper confidence level (84\%) minus the lower confidence level (16\%) for the atmospheric parameters of the sample \citep{2023A&A...674A..26C,2023A&A...674A..28F,2023A&A...674A..29R}. For the other surveys, the typical uncertainty is defined as the median value of the errors in the atmospheric parameters of the sample \citep{2022ApJS..259...35A,2024arXiv240919858B}. N refers to the number of common stars.}
\end{table}

\subsection{Method}\label{sec:method}

To evaluate the quality of the stellar parameter values in {\it Gaia} DR3, LAMOST-LRS DR11, APOGEE DR17, and GALAH DR4, we adopted members in 130 open clusters with well-defined main sequences from the OCSN catalog as standards. While we adopted the metallicity and age values of open clusters from \citetalias{2023ApJS..265...12Q}, we acknowledge that these reference metallicities may have errors that could introduce potential biases into our analysis. For this work, we assume that these are the true values. We estimated the atmospheric parameters of member stars through the isochrone fitting approach. Then we compared the estimated parameters to those from different catalogs. The steps are listed below:

\begin{enumerate}

\item[(1)] We prepared the theoretical nonrotating isochrones for this analysis. It is important to note that the use of nonrotating models may limit the accuracy for early-type stars, particularly those of spectral types O, B, and A, where rotation plays a significant evolutionary role. Based on the cluster age and metallicity parameters provided by \citetalias{2023ApJS..265...12Q}, we obtained the PARSEC isochrones \citep{2017ApJ...835...77M} with the {\it Gaia} photometric system \citep{2021yCat..36490003R} from CMD 3.7\footnote{\url{http://stev.oapd.inaf.it/cmd}} for each cluster. Furthermore, for each isochrone, we interpolated the parameters (including $G_{\rm iso}$, $BP_{\rm iso}$, $RP_{\rm iso}$, $T_{\rm eff\_iso}$, $\log g_{\rm\_iso}$) on a mass grid with a step size of $0.0005\,\rm M_{\odot}$.

\item[(2)] We obtained the intrinsic color and absolute magnitude. Based on the distance modulus $(m-M)_0$ and reddening value $E(B-V)$ given by \citetalias{2023ApJS..265...12Q} for each cluster, we used $A_{\rm G}=2.74 \times E(B-V)$ and $E(BP-RP)=1.339 \times E(B-V)$ \citep{2018MNRAS.479L.102C,2019A&A...624A..34Z} to calculate the absolute magnitude, $M_{\rm G}=G_{\rm obs}- (m-M)_0-A_{\rm G}$, and the intrinsic color, $(BP-RP)_{0}=(BP-RP)_{\rm obs}-E(BP-RP)$, of each member star. It should be noted that the extinction coefficient here is applicable to stars in the temperature range of [5250, 7000]\,\rm K. However, it has been applied for all stars, and readers are advised to pay attention to this point (for more discussions, please refer to Sect.~\ref{sec:D5}).

\item[(3)] We estimated the theoretical atmospheric parameters. On the CMD of each cluster, we matched each member to the theoretical point in the isochrone with the minimum distance. The minimum distance is defined as the minimum difference between the observed and theoretical values of $\sqrt{\Delta(BP-RP)_0^2+\Delta M_{\rm G}^2}$. Then, we obtained theoretical $T_{\rm eff\_iso}$ and $\log g_{\rm\_ iso}$ from the isochrone for each member star. At the same time we also considered the uncertainty due to observational errors. We provide these estimated theoretical atmospheric parameters and the uncertainties in Appendix~\ref{Appendix}.

\item[(4)] We calculated the deviations between observational and theoretical atmospheric parameters. Adopting theoretical values as reference values, we separately calculated the differences between the crossmatched atmospheric parameters and the reference values, which are $\Delta T_{\rm eff\_surveys}=T_{\rm eff\_surveys}-T_{\rm eff\_iso}$, $\Delta$log $g_{\rm \_surveys}=$ log $g_{\rm\_surveys}$ $-$ log $g_{\rm\_iso}$. Then, the data quality of these catalogs were evaluated.

\end{enumerate}

\begin{figure*}[!htbp]
\includegraphics[width=0.99\textwidth]{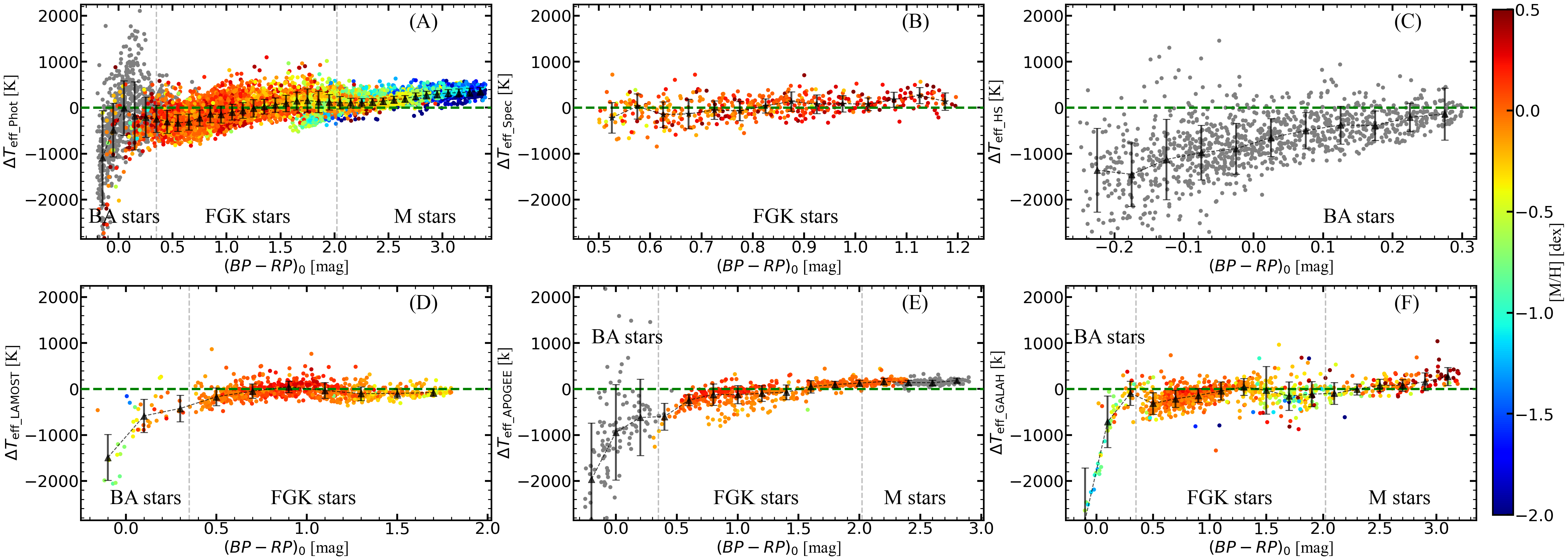}
\caption{$\Delta T_{\rm eff}$ vs. $(BP-RP)_{0}$. The $\Delta T_{\rm eff\_Phot/Spec/HS/LAMOST/APOGEE/GALAH}$ is defined as $T_{\rm eff\_Phot/Spec/HS/LAMOST/APOGEE/GALAH}$ $-$ $T_{\rm eff\_iso}$. The black triangles and error bars indicate the median values and corresponding dispersions within different color bins. The vertical dashed gray lines are the cutoffs between different stellar types. The rainbow color of the points represents the metallicity from individual catalogs with a color bar on the right. The metallicity of the GSP-Phot has been calibrated with the calibration relation from \citet{2023A&A...674A..27A}. Gray points refer to the samples without observational metallicity.}
\label{Fig3}
\end{figure*}

\begin{figure*}[!htbp]
\includegraphics[width=0.99\textwidth]{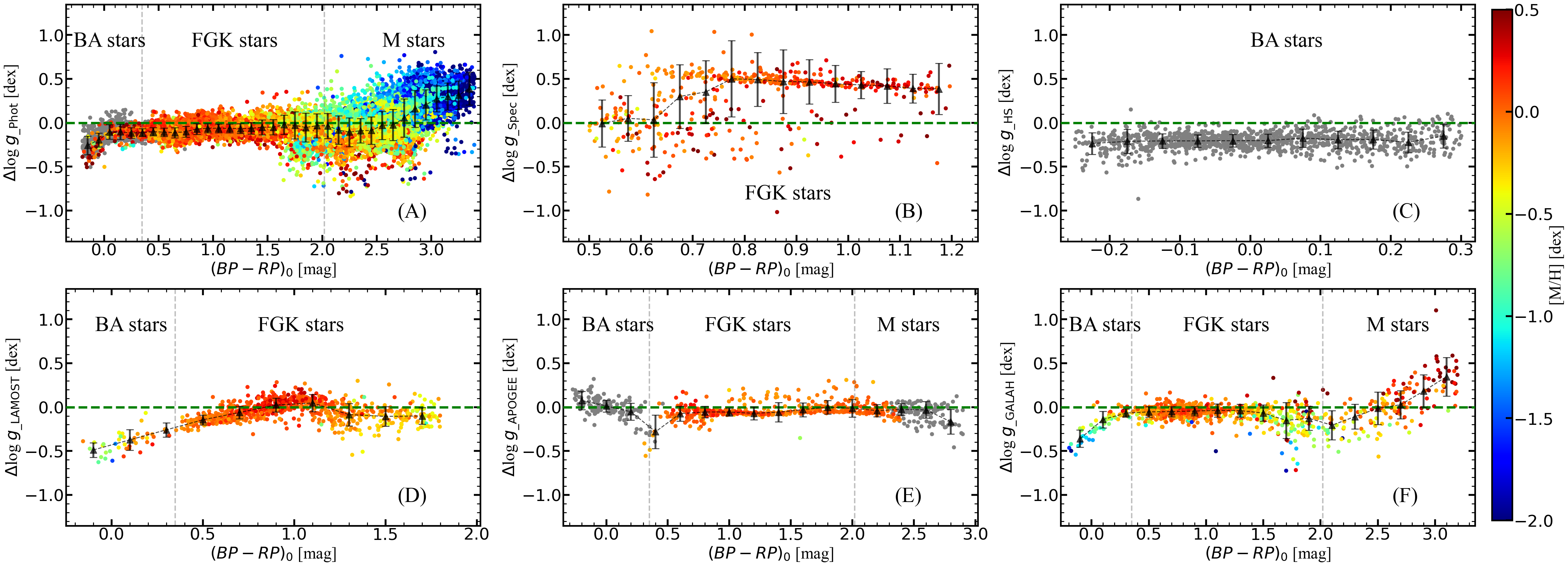}
\caption{Same as Fig.~\ref{Fig3}, but for $\Delta$log $g$. The $\Delta$log $g_{\rm\_ Phot/Spec/HS/LAMOST/APOGEE/GALAH}$ is defined as and log $g_{\rm \_Phot/Spec/HS/LAMOST/APOGEE/GALAH}$ $-$ log $g_{\rm \_iso}$.}
\label{Fig4}
\end{figure*}

\section{Validating atmospheric parameters}\label{sec:Vali-AP}

We calculated the median and dispersion values for the differences between the theoretical and observational parameters for the selected cluster member stars. Then we evaluated the atmospheric parameters provided by {\it Gaia} GSP-Phot, GSP-Spec, ESP-HS, LAMOST, APOGEE, and GALAH, and summarized them in Table~\ref{tab:gaia}. The crossmatched catalogs were given in Col.1 (`Catalogs'). Based on the deviation variations in Fig.~\ref{Fig3} and Fig.~\ref{Fig4}, we roughly divided the member sample into different stellar type groups (Col. 2, “Stellar type”) based on the range of intrinsic colors, and calculated the corresponding median deviations (Col. 3 and 5, “Med($\Delta T_{\rm eff}$)” and “Med($\Delta$log $g$)”) and the dispersions (Col. 4 and 6, “$\sigma(\Delta T_{\rm eff})$” and “$\sigma(\Delta \log g$)”) of stars.

\begin{table*}[!htbp]
\renewcommand{\arraystretch}{1.3}
\caption{Summary of the median values and the dispersions of $\Delta T_{\rm eff}$ and $\Delta$log $g$ in different type stars for GSP-Phot, GSP-Spec, ESP-HS, LAMOST-LRS DR11, APOGEE DR17, and GALAH DR4.}\label{tab:gaia}
\centering
\large
\scalebox{1}{
\begin{tabular}{lccccc}
\hline\hline
Catalog & Stellar Type & Med($\Delta T_{\rm eff}$) & $\sigma(\Delta T_{\rm eff})$ & Med($\Delta \log g$) & $\sigma(\Delta \log g)$  \\ 
 &  & (K) & (K) & (dex) & (dex) \\ 
\hline
\multirow{3}{*}{GSP-Phot} 
    & BA  & $-$314 & 775  & $-$0.13 & 0.1  \\ 
    & FGK & $-$102 & 243  & $-$0.08 & 0.09  \\ 
    & M   & 197  & 122  &  0.02 & 0.22 \\ 
\hline
GSP-Spec & FGK & 35 & 242 & 0.43 & 0.33 \\ 
\hline
ESP-HS & BA & $-$681 & 704 & $-$0.2 & 0.1  \\ 
\hline
\multirow{2}{*}{LAMOST} 
    & BA  & $-$694  & 581  & $-$0.4 & 0.12  \\ 
    & FGK & $-$59   & 161  & $-$0.05 & 0.11  \\ 
\hline
\multirow{3}{*}{APOGEE} 
    & BA  & $-$1043 & 1229 &  0.01 & 0.13  \\ 
    & FGK & $-$62   & 220  & $-$0.05 & 0.09  \\ 
    & M   & 137   & 61   & $-$0.03 & 0.09  \\ 
\hline
\multirow{3}{*}{GALAH} 
    & BA  & $-$708  & 1383 & $-$0.14 & 0.14  \\ 
    & FGK & $-$132  & 258  & $-$0.05 & 0.1  \\ 
    & M   & 81    & 203  &  0 & 0.26  \\ 
\hline
\end{tabular}}
\end{table*}

Fig.~\ref{Fig3} and Fig.~\ref{Fig4} display the parameter deviation ($\Delta T_{\rm eff}$ and $\Delta \log g$) distributions for different spectroscopic surveys within different intrinsic color ranges. For the GSP-Phot sample, in the intrinsic color range of [$-$0.2, 3.4]\,\rm mag, we calculated the median and dispersion values with a color step size of $0.1\,\rm mag$. For the GSP-Spec sample, we computed the median and dispersion values in the intrinsic colors range of [0.5, 1.2]\,\rm mag with a step of $0.05\,\rm mag$. For the ESP-HS samples, we made similar calculations in the intrinsic color range of [$-$0.25, 0.3]\,\rm mag with a step of 0.05\,\rm mag. For the LAMOST sample, we analyzed the parameter deviations in the intrinsic color range of [$-$0.2, 1.8]\,\rm mag with a step of 0.2\,\rm mag. For the APOGEE sample, we evaluated the parameter deviations in the intrinsic color range of [$-$0.3, 2.9]\,\rm mag with a step of 0.2\,\rm mag. Within the intrinsic color range of [$-$0.2, 3.2]\,\rm mag, we performed a similar estimation for the GALAH sample with a step of 0.2\,\rm mag.

The metallicity of each catalog is also shown in Fig.~\ref{Fig3} and Fig.~\ref{Fig4}. The metallicity of the GSP-Phot has been calibrated with the calibration relation from \citet{2023A&A...674A..27A}. We find that the metallicity of the B, A, and K-type stars of LAMOST-LRS DR11, GALAH DR4 seems lower than that of the G, F-type stars. In addition, there is an underestimation of the metallicity of M-type stars in GSP-Phot.

\subsection[Teff, log g from GSP-Phot]{$T_{\rm eff}$, $\log g$ from GSP-Phot}\label{sec:gspphot}

\begin{figure}[!htbp]
\includegraphics[width=0.49\textwidth]{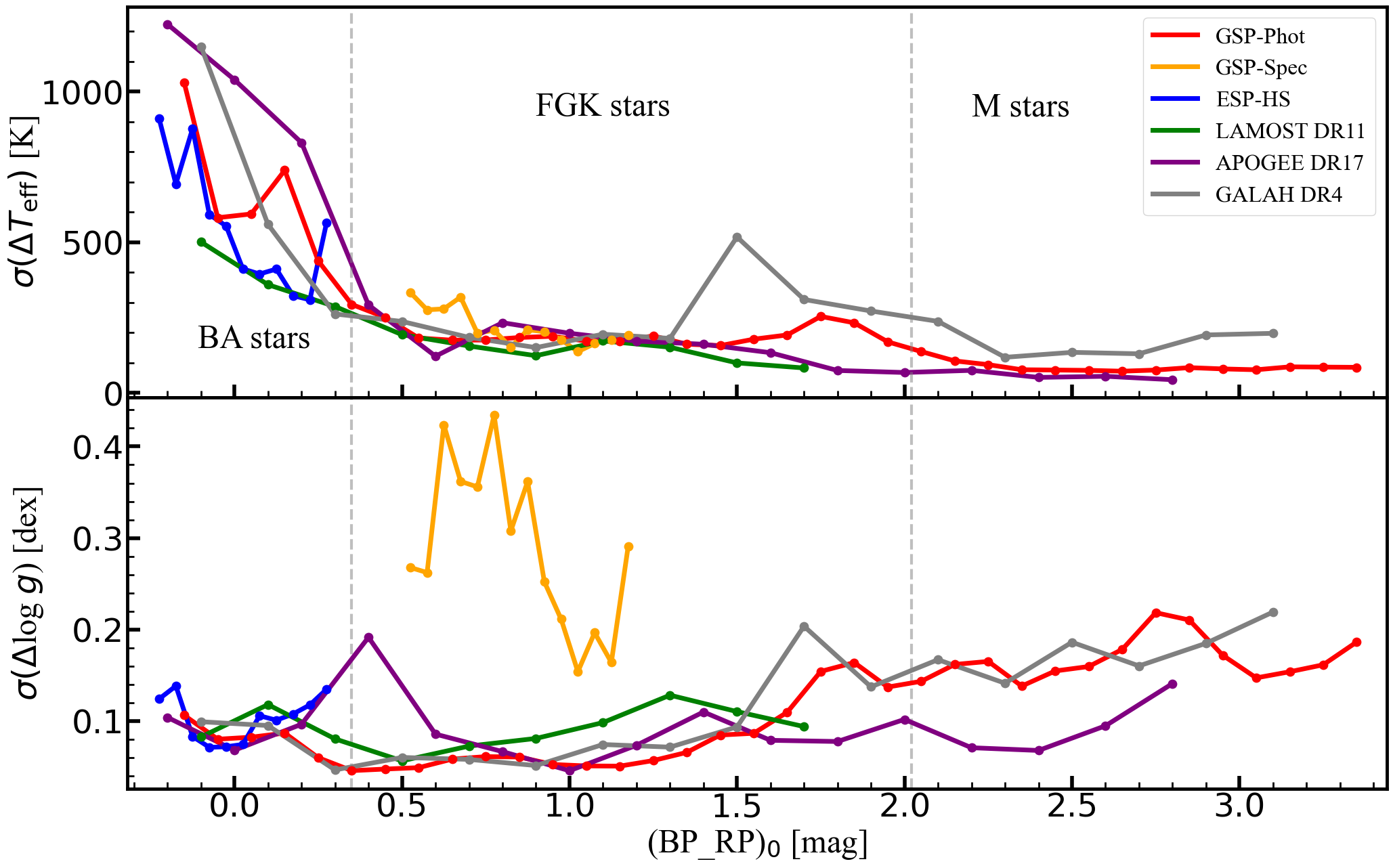}
\caption{Top panel: Dispersion of $\Delta T_{\rm eff}$ vs. $(BP-RP)_{0}$. Bottom panel: Dispersion of $\Delta$log $g$ vs. $(BP-RP)_{0}$. The red, orange, blue, green, purple, and gray lines represent the dispersion of $\Delta T_{\rm eff}$ and $\Delta$log $g$ of GSP-Phot, GSP-Spec, ESP-HS, LAMOST-LRS DR11, APOGEE DR17, and GALAH DR4, respectively.}
\label{dispersion}
\end{figure}

Panel (A) in Fig.~\ref{Fig3} and Fig.~\ref{Fig4} presents the distribution of $\Delta T_{\rm eff\_Phot}$ and $\Delta$log $g_{\rm\_Phot}$ versus the intrinsic color. The median values for $\Delta T_{\rm eff\_Phot}$ and $\Delta \log{g}_{\rm\_Phot}$ of all member stars are $101\,\rm K$ and $-0.06\,\rm dex$, respectively, while their MAD values are $232\,\rm K$ and $0.14\,\rm dex$. It is noted that the MAD values of the difference ($\Delta T_{\rm eff\_Phot}$ and $\Delta \log{g}_{\rm\_Phot}$) in our sample are lower than those ($400\,\rm K$ and $0.22\,\rm dex$) provided by~\citetalias{2023A&A...674A..28F}. Moreover, as is shown in Fig.~\ref{dispersion} (red lines), the median dispersions of $\Delta T_{\rm eff\_Phot}$ and $\Delta \log{g}_{\rm\_Phot}$ are 174~K and 0.11~dex for all the selected sample stars, which are smaller than those from \citet{2023A&A...674A..27A} (see their Table 1 and Table 2). This may be because the selected comparison sample in this work is the high-quality open clusters with well-defined main sequences, and stars with a binary mass ratio larger than 0.5 were excluded. We discuss the variations of $T_{\rm eff}$ and $\log g$ deviations for different types of stars in color bins, which are shown below:

\begin{enumerate}
\item[(1)] B and A-type stars: both $\Delta T_{\rm eff\_Phot}$ and the corresponding dispersions show significant variations with a median deviation of $-233\,\rm K$ and a median dispersion of $593\,\rm K$. The median value of $\Delta$log $g_{\rm\_Phot}$ and the corresponding median dispersion are $-0.11\,\rm dex$ $0.08\,\rm dex$.

\item[(2)] F, G, and K-type stars: $\Delta T_{\rm eff\_Phot}$ shows a median value of $-40\,\rm K$ and a median dispersion of $186\,\rm K$. When the stellar color changed from F-type to K-type, $\Delta T_{\rm eff\_Phot}$ gradually increased from $-320\,\rm K$ to $150\,\rm K$, which is generally in agreement with \citetalias{2023A&A...674A..28F}.
The log $g_{\rm\_Phot}$ of the F, G, and K-type stars is mainly underestimated with a median value of $-0.06\,\rm dex$, and the corresponding dispersion is $0.06\,\rm dex$, which is similar to the result of \citetalias{2023A&A...677A.162B}, while \citetalias{2023A&A...674A..28F} obtained the opposite conclusion. Moreover, we found a bifurcation structure of deviations for K-type stars.

\item[(3)] M-type stars: the $\Delta T_{\rm eff\_Phot}$ increases from $100\,\rm K$ to $350\,\rm K$ as the color reddens, while the dispersions are relatively stable. As M-type stars redden, the deviation of $\log g$ increases from $-0.1\,\rm dex$ to $0.39\,\rm dex$. The median values of $\Delta T_{\rm eff\_Phot}$ and $\Delta$log $g_{\rm\_Phot}$ are $213\,\rm K$ and $0.01\,\rm dex$, and the median dispersions are $86\,\rm K$ and $0.16\,\rm dex$.

\end{enumerate}

\begin{figure}[!htbp]
\centering
\includegraphics[width=0.49\textwidth]{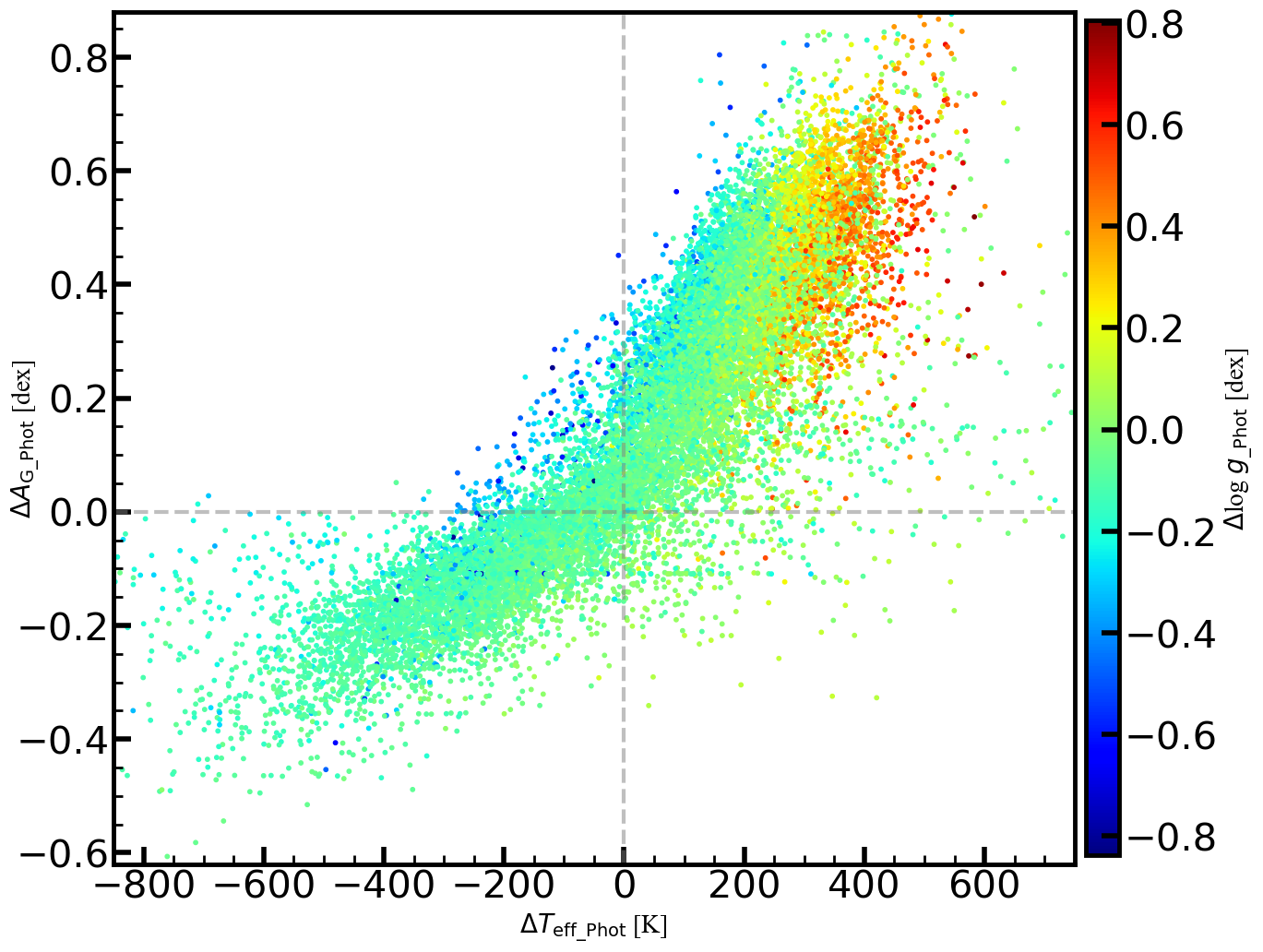}
\caption{$\Delta T_{\rm eff\_Phot}$ vs. $\Delta A_{\rm G\_Phot}$. $\Delta A_{\rm G\_Phot}$ is defined as $A_{\rm G\_Phot}$ $-$ $A_{\rm G\_iso}$. The color of the points represents $\Delta$log $g_{\rm\_Phot}$ with a color bar on the right.}
\label{Fig6}
\end{figure}

\cite{2023A&A...674A..27A} point out that there is a degeneracy problem between $T_{\rm eff} $ and extinction in the GSP-Phot, which means that an observed red star could be a hot star reddened owing to dust extinction on the line of sight, or a truly cool star. This makes it difficult for GSP-Phot to differentiate the two cases. Fig.~\ref{Fig6} displays the distribution of $\Delta A_{\rm G\_{\rm Phot}}$ as a function of $\Delta T_{\rm eff\_Phot}$ in our sample, with the color of each point representing the $\Delta$ log $g_{\rm\_Phot}$. Similarly, our result shows that the $T_{\rm eff}$ strongly degenerates with extinction, which may be responsible for the deviation of GSP-Phot's atmospheric parameters from the theoretical values. The linear relation demonstrates that the temperature and extinction in the $G$ band for most stars are simultaneously overestimated or underestimated. For stars with severely overestimated temperatures and extinctions, the GSP-Phot also overestimates their log $g$, and vice versa. 

\begin{figure}[!htbp]
\includegraphics[width=0.48\textwidth]{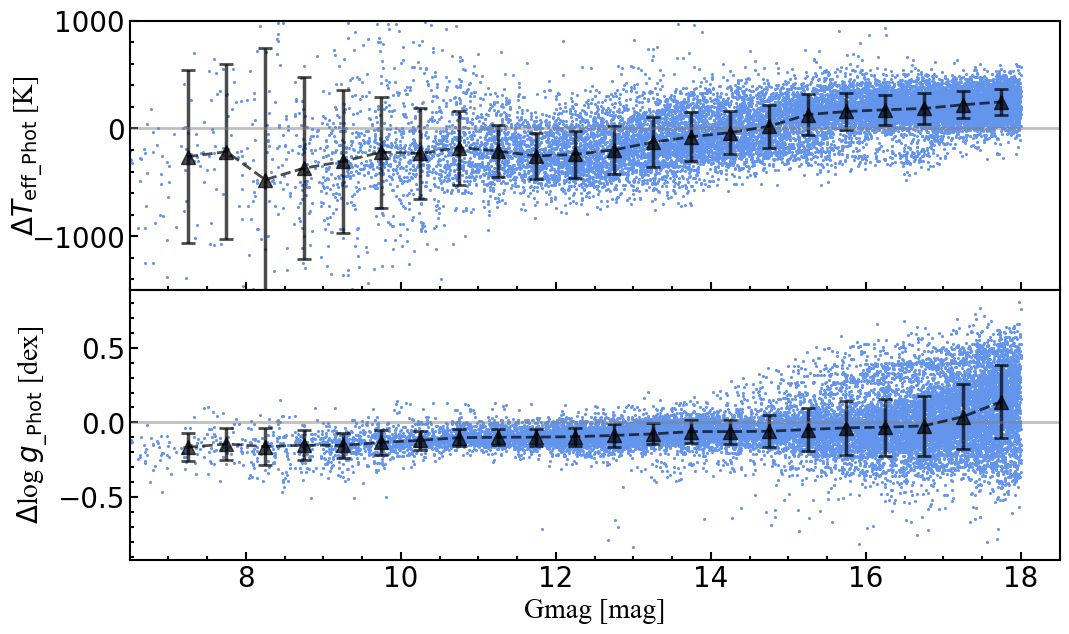}
\caption{Top panel: $\Delta T_{\rm eff\_Phot}$ vs. Gmag. Bottom panel: $\Delta$log $g\rm_{\_Phot}$ vs. Gamg. The black triangles and error bars indicate the median values and corresponding dispersions within different color bins.}
\label{Gamg_dev}
\end{figure}

In addition, Fig.~\ref{Gamg_dev} shows the dependence of $\Delta T_{\rm eff\_Phot}$ and $\Delta$log $g\rm_{\_Phot}$ on $G$ magnitude in the range of [6, 18]\,\rm mag with a step of 0.5~mag. We found that $\Delta T_{\rm eff\_Phot}$ varies from 240\,\rm K for fainter stars to $-460$\,\rm K for brighter stars, $\Delta$log $g\rm_{\_Phot}$ changes from 0.14\,\rm dex to $-0.16$\,\rm dex as stars become brighter, and the dispersion of $\Delta$log $g\rm_{\_Phot}$ is larger for the fainter stars.

\subsection[Teff, log g from GSP-Spec]{$T_{\rm eff}$, $\log g$ from GSP-Spec}\label{sec:gspspec}

The common GSP-Spec sources are mainly F and G-type stars, and some K-type stars. The distributions of $\Delta T_{\rm eff\_Spec}$ and $\Delta$log $g_{\rm\_Spec}$ are shown in panel (B) of Fig.~\ref{Fig3} and Fig.~\ref{Fig4}. For these stars, the median value of $\Delta T_{\rm eff\_Spec}$ is $60\,\rm K$, which is relatively small, while the dispersion is $202\,\rm K$.
The median value of $\Delta$log $g_{\rm\_Spec}$ is $0.4\,\rm dex$ and the dispersion is $0.3\,\rm dex$. We noted a clear linear relation of the $\Delta$log $g_{\rm\_Spec}$ in the color range of $[0.6, 1.2]\,\rm mag$, with a decrease from $0.6\,\rm dex$ to $0.3\,\rm dex$.
In addition, we calibrated the log $g_{\rm\_Spec}$ according to the polynomial provided by~\cite{2023A&A...674A..29R}. Although this linear structure still exists after calibration, the polynomial reduces the median deviation of log $g_{\rm\_Spec}$ from $0.44\,\rm dex$ to $0.31\,\rm dex$. In Fig.~\ref{dispersion}, $\Delta$log $g_{\rm\_Spec}$ (orange line) shows large dispersions. The median dispersions of $\Delta T_{\rm eff\_Spec}$ and $\Delta$log $g_{\rm\_Spec}$ in our work are larger than those in~\cite{2023A&A...674A..29R}, because they employed a strict data quality control. In their sample, they selected the samples where the first 13 quality flags of gspspec\_flags are all equal to 0, while in our sample, only the first six of these flags are equal to 0.

\subsection[Teff, log g from ESP-HS]{$T_{\rm eff}$, $\log g$ from ESP-HS}\label{sec:esphs}

It should be pointed out that solar metallicity was assumed for all the sources in the ESP-HS module \citepalias{2023A&A...674A..28F}. Most selected ESP-HS sources are B and A-type stars, and the distributions of $\Delta T_{\rm eff\_HS}$ and $\Delta$log $g_{\rm\_HS}$ versus intrinsic color are shown in panel (C) of Fig.~\ref{Fig3} and Fig.~\ref{Fig4}. The median value of $\Delta T_{\rm eff\_HS}$ is $-653\,\rm K$. There is a clear decrease in the underestimation of $T_{\rm eff\_HS}$ from B-type to A-type stars, from $1450\,\rm K$ to $137\,\rm K$. Meanwhile, the $\Delta T_{\rm eff\_HS}$ dispersion of B-type stars is generally larger than that of A-type stars. ESP-HS underestimates the log $g_{\rm\_HS}$ with a relatively stable deviation of $-0.2\,\rm dex$ and a dispersion of $0.1\,\rm dex$. Fig.~\ref{dispersion} displays the parameter dispersion distribution (blue lines). The dispersion of $\Delta T_{\rm eff\_HS}$ in this work is similar to the result of \citetalias{2023A&A...674A..28F}. But the dispersions of $\Delta T_{\rm eff\_HS}$ are smaller than those in \citetalias{2023A&A...674A..28F}, with a dispersion of 0.2~dex for A-type stars to 0.4~dex for O-type stars.

\subsection[Teff, log g from LAMOST-LRS DR11]{$T_{\rm eff}$, $\log g$ from LAMOST-LRS DR11}\label{sec:lamost}

As is shown in panel (D) of Fig.~\ref{Fig3} and Fig.~\ref{Fig4}, the common member stars with LAMSOT-LRS DR11 are mainly F, G, and K-type stars and a few B and A-type stars. Both $\Delta T_{\rm eff\_LAMOST}$ and $\Delta$log $g_{\rm\_LAMOST}$ show an increasing trend from B-type to G-type stars, followed by a minor decline from G-type to K-type stars. The median $\Delta T_{\rm eff\_LAMOST}$ and $\Delta$log $g_{\rm\_LAMOST}$ are $-85$\,\rm K and $-0.09$\,\rm dex for F, G, and K-type stars. The corresponding median dispersions are of 157\,\rm K and 0.09\,\rm dex. 
The deviations are more significant for B and A-type stars with a median value of $-579$\,\rm K in $\Delta T_{\rm eff\_LAMOST}$ and $-0.37$\,\rm dex in $\Delta$log $g_{\rm\_LAMOST}$, respectively. The corresponding median dispersions are 371\,\rm K and 0.08\,\rm dex. As is shown in Fig.~\ref{dispersion} (green lines), the $\Delta T_{\rm eff\_LAMOST}$ dispersion decreases as stars become redder, and the $\Delta$log $g_{\rm\_LAMOST}$ dispersions are smaller than 0.15\,\rm dex with minimal variation.

\subsection[Teff, log g from APOGEE DR17]{$T_{\rm eff}$, $\log g$ from APOGEE DR17}\label{sec:apogee}

Panel (E) of Fig.~\ref{Fig3} and Fig.~\ref{Fig4} shows the deviation of the $T_{\rm eff}$, log $g$ of APOGEE DR17 from the theoretical parameters, respectively. 
The $\Delta T_{\rm eff\_APOGEE}$ increases as stars get redder in the color range of [$-$0.2, 2.0]\,\rm mag, from the median $\Delta T_{\rm eff\_APOGEE}$ of $-942$\,\rm K for B and A-type stars to $-101$\,\rm K for F, G, and K-type stars. 
For M-type stars, we determined a median $\Delta T_{\rm eff\_APOGEE}$ offset of 145\,\rm K with corresponding median dispersions of 57\,\rm K. We also found a median $\Delta$log $g_{\rm\_APOGEE}$ offset $-0.06$\,\rm dex of for late F, G, K, and M-type stars, which indicates that the observational data is quite consistent with the theoretical model.
However, the $\Delta$log $g_{\rm\_APOGEE}$ shows a declining trend for B and A-type stars. As is shown in Fig.~\ref{dispersion} (purple lines), the $\Delta T_{\rm eff\_APOGEE}$ dispersion decreases as stars become redder, and the $\Delta$log $g_{\rm\_APOGEE}$ dispersions exhibit noticeable variation.

\subsection[Teff, log g from GALAH DR4]{$T_{\rm eff}$, $\log g$ from GALAH DR4}\label{sec:galah}

The deviations of the atmospheric parameters of GALAH DR4 from the theoretical values are shown in panel (F) of Fig.~\ref{Fig3} and Fig.~\ref{Fig4}. The $\Delta T_{\rm eff\_GALAH}$ shows an increasing trend for F, G-type stars, a decline feature for K-type stars, and then an increasing trend for M-type stars.
We can see a $\Delta$log $g_{\rm\_GALAH}$ offset of $-$0.05\,\rm dex for F and G-type stars and an increasing trend for K and M-type stars. Due to the limited sample number, both $\Delta T_{\rm eff\_GALAH}$ and $\Delta$log $g_{\rm\_GALAH}$ exhibit increasing features for B and A-type stars. As is shown in Fig.~\ref{dispersion} (gray lines), for B, A, F, and G-type stars, the $\Delta T_{\rm eff\_GALAH}$ dispersion decreases as stars become cooler, and the $\Delta$log $g_{\rm\_GALAH}$ dispersions are smaller than 0.15\,\rm dex with slight variation. Then the dispersions of both parameters increase for K and M-type stars.

Overall, the parameter deviations and dispersions of different spectroscopic surveys exhibit different variations. For all the above surveys, the parameter deviations of F, G, and K-type stars are smaller than those of B, A, and M-type stars. 
Specifically, for B and A-type stars, the $T_{\rm eff}$ deviations of {\it Gaia} data (including GSP-Phot and ESP-HS) are notably lower compared to other surveys. Conversely, for M-type stars, the $T_{\rm eff}$ deviations of GSP-Phot are significantly higher than those observed in APOGEE and GALAH. Among F, G, and K-type stars, the $T_{\rm eff}$ deviations are the largest for GALAH, followed by GSP-Phot, APOGEE, LAMOST, and finally GSP-Spec.
As is shown in the top panel of Fig.~\ref{dispersion}, the $\Delta T_{\rm eff}$ dispersions decrease as stars become cooler for most samples, and the $\Delta T_{\rm eff}$ dispersions of K-type stars from GSP-Phot and GALAH are larger than the ones for the F and G-type stars. In the bottom panel of Fig.~\ref{dispersion}, we can see that the $\Delta \log g$ dispersions of K and M-type stars are larger than the ones for the B, A, F, and G-type stars for most samples.
However, the $\Delta \log g$ dispersions of GSP-Spec samples are significantly larger than those in other surveys.

\section{Discussion}\label{sec:D}

\subsection{The influence of stellar rotation}\label{sec:D1}

Stellar rotation influences the atmospheric parameter derivation across all surveys. This is particularly evident in the systematic offsets observed for B and A-type stars, as has been reported consistently by multiple catalogs. We have found that the B and A-type stars of all surveys both $T_{\rm eff}$ and log $g$ are underestimated. This situation is likely associated with our failure to utilize the rotational isochrones. We employ the atmospheric parameters furnished by the rotational isochrones as a theoretical benchmark to explore how stellar rotation impacts our results.

\begin{figure}[!htbp]
\includegraphics[width=0.48\textwidth]{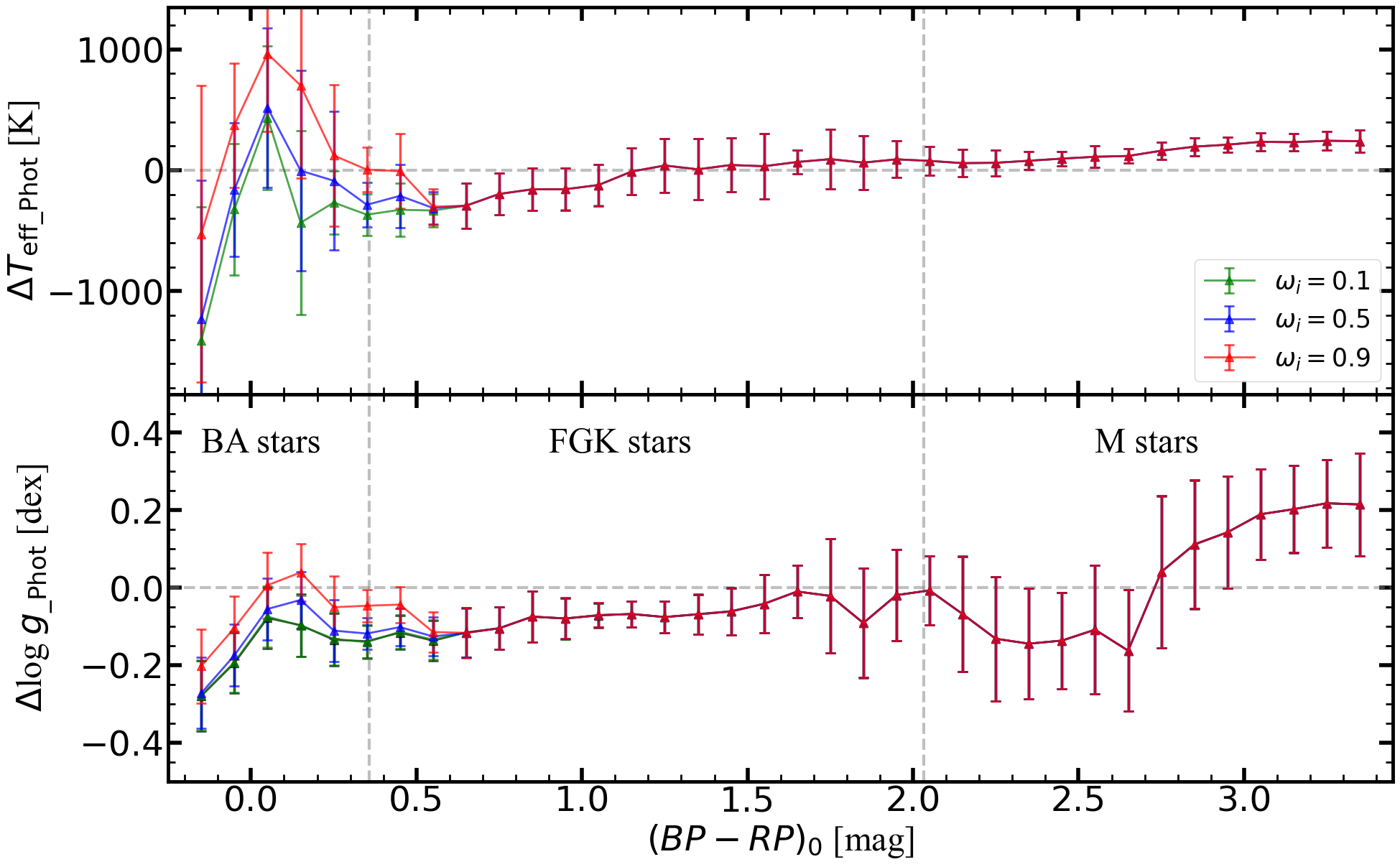}
\caption{Deviation between the $T_{\rm eff}$ (upper panel) and log $g$ (lower panel) obtained by GSP-Phot and the theoretical values of the isochrone models with different degrees of stellar rotation. The green, yellow, and red lines represent, respectively, the deviations from the theoretical isochrones with rotational angular velocities of $\omega_i$ = 0.1, 0.5, and 0.9. The error bars denote the degree of dispersion in each color bin.}
\label{fig:D1}
\end{figure}

We have selected six open clusters (OCSN\_77, OCSN\_90, OCSN\_212, OCSN\_218, OCSN\_219, and OCSN\_259) from our entire sample as test cases. For each cluster, we derived theoretical atmospheric parameter estimates using isochrones with specific angular rotation rates ($\omega _{i}=0.1, 0.5, 0.9$). Fig.~\ref{fig:D1} presents the distribution of deviations between GSP-Phot atmospheric parameters and those derived from rotational isochrone models. Here, when we analyze the impact of rotation on GSP-Phot results, it should be understood as a case study representative of broader methodological issues affecting all datasets.

The lower panel of Fig.~\ref{fig:D1} shows that the isochrone models considering stellar rotation can reduce the degree of deviation between the log $g_{\rm\_Phot}$ of the B, A, and early F-type stars obtained by GSP-Phot and the theoretical values. In other words, the theoretical values of the log $g_{\_iso}$ obtained from the isochrone models with stellar rotation are smaller than those from the isochrone models without stellar rotation. As is shown in the upper panel of Fig.~\ref{fig:D1}, the theoretical values of the $T_{\rm eff\_iso}$ obtained from the isochrone model with stellar rotation are also smaller than that from the isochrone model without stellar rotation. We have summarized the typical deviations and dispersions of the $T_{\rm eff}$ and log $g$ of B and A-type stars under different rotational isochrones, as is shown in Table~\ref{tab:D1}.

However, for some A-type stars, the isochrone with a much higher rotation ($\omega_i=0.9$) will yield an even larger deviation of the $T_{\rm eff}$ (as is indicated by the red line). By contrast, stellar rotation has almost no impact on late F, G, K, and M-type stars. While we use GSP-Phot as a specific example to illustrate these effects, similar discrepancies are observed across all surveys. This suggests that the source of the offsets could lie in limitations of the reference temperatures or methodologies rather than specific pipelines.

\begin{table}[!htbp]
\renewcommand{\arraystretch}{1.3}
\caption{Summary of the median values and the dispersions of $\Delta T_{\rm eff}$ and $\Delta$log $g$ in B and A-type stars for GSP-Phot from different rotational isochrones models.}\label{tab:D1}
\centering
\large
\scalebox{.98}{
\begin{tabular}{lcccc}
\hline\hline
\multirow{2}{*}{$\omega_i$} 
 & Med($\Delta T_{\rm eff}$) & $\sigma(\Delta T_{\rm eff})$ & Med($\Delta \log g$) & $\sigma(\Delta \log g)$  \\ 
 &  (K) & (K) & (dex) & (dex) \\ 
\hline
0.1  & $-$423 & 912 & $-$0.16 & 0.11 \\ 
\hline
0.5  & $-$275 & 957 & $-$0.13 & 0.11  \\ 
\hline
0.9  & 104 & 915 & $-$0.07 & 0.12  \\ 
\hline
\end{tabular}}
\end{table}

\subsection{Dispersion at the low-mass end}\label{sec:D2}

\begin{figure}[!htbp]
\includegraphics[width=0.48\textwidth]{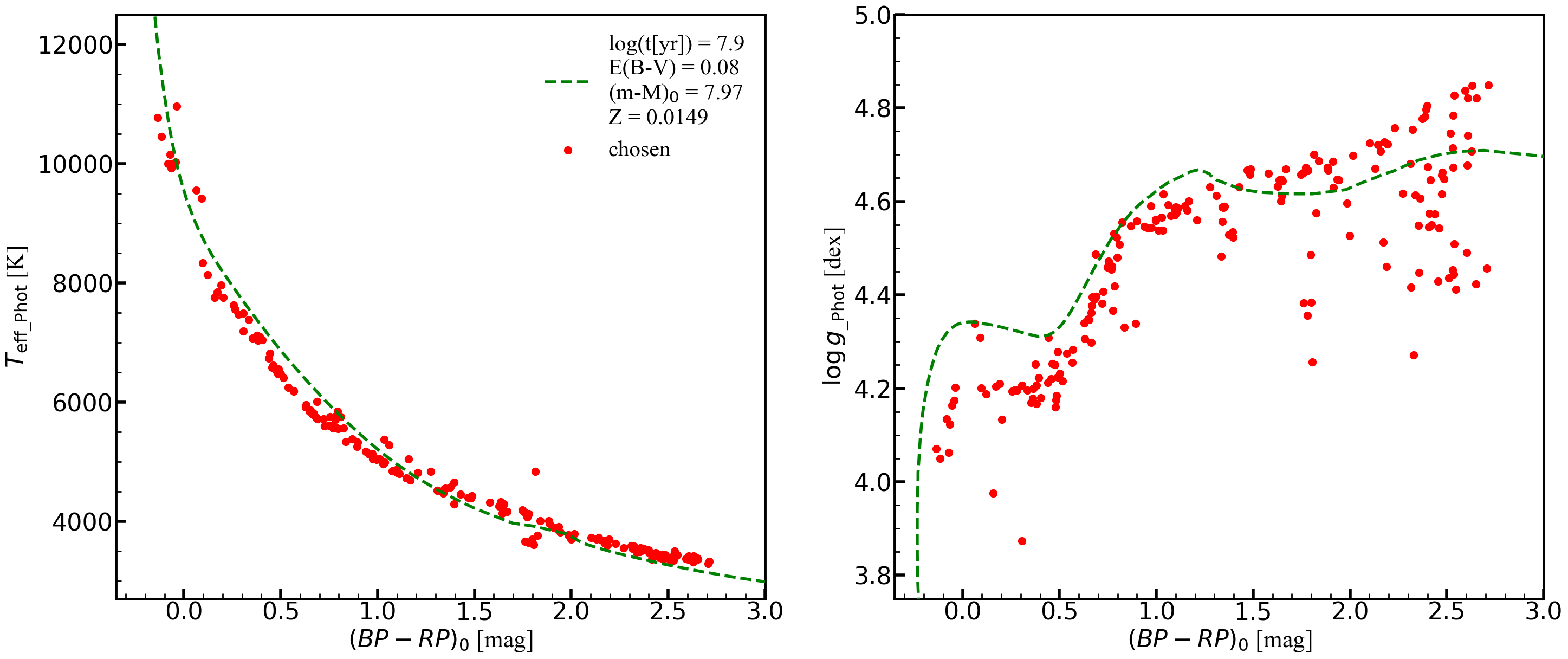}
\caption{Distribution of $T_{\rm eff}$ (left) and log $g$ (right) from GSP-Phot for the retained member stars of OCSN\_259 as a function of intrinsic color. The dashed green line represents the isochrone. Red dots indicate the retained member stars.}
\label{fig:D2}
\end{figure}

The pronounced dispersion observed among low-mass stars in the CMD potentially impacts atmospheric parameter analyses across all surveys. To discuss whether the dispersion here is comparable to that of the atmospheric parameters, we presented the distribution of the observed $T_{\rm eff\_Phot}$ and log $g_{\rm\_Phot}$ of GSP-Phot with respect to the intrinsic colors for the member stars retained in Fig.~\ref{Fig2}, as is shown in Fig.~\ref{fig:D2}. Consequently, the analysis of how the low-mass-end dispersion impacts GSP-Phot results also serves as a case study to illustrate the influence of the dispersion at the low-mass end that is brought to all the other survey catalogs.

For the low-mass end, such as the member stars with intrinsic colors around $2.5\,\rm mag$ in Fig.~\ref{Fig2}, these stars exhibit a relatively significant deviation from the isochrones and also display a relatively large dispersion. In the same color region of the left panel of Fig.~\ref{fig:D2}, the $T_{\rm eff\_Phot}$ show a systematic overestimation, and the dispersion is not significant. However, as is shown in the right panel of Fig.~\ref{fig:D2}, for the log $g_{\rm\_Phot}$, there are significant deviations and dispersion in the observed values. Moreover, at the blue end of $(BP-RP)_0$, there is also a significant underestimation and dispersion of the $\log g_{\rm\_Phot}$.

Although there is a relatively large dispersion in observed color and magnitude in the member stars at the low-mass end in the CMD, the observed values of their $T_{\rm eff\_Phot}$ exhibit a smaller dispersion. The dispersion of the observed values of the log $g_{\rm\_Phot}$ is significantly larger. This means that the measurement of the $T_{\rm eff}$ at the low-mass end is relatively reliable, while the measurement error of the $\log g$ at the low-mass end is relatively large, showing a significant dispersion.

\subsection{The influence of the extinction law}\label{sec:D5}

As is well known, the extinction coefficient depends on the $T_{\rm eff}$ and $\log g$ of the stars \citep{2010A&A...523A..48J}. The fixed extinction coefficient we used could potentially affect all our results. To this end, we utilized the extinction law of {\it Gaia} (E)DR3\footnote{\hbox{\url{https://www.cosmos.esa.int/web/gaia/edr3-extinction-law}}}\citep{2021A&A...649A...3R} to calculate the extinction coefficients of member stars at different temperatures, aiming to explore the influence of different extinction coefficients on our results. We took the parameters from GSP-Phot as an example for discussion.

\begin{figure}[!htbp]
\includegraphics[width=0.48\textwidth]{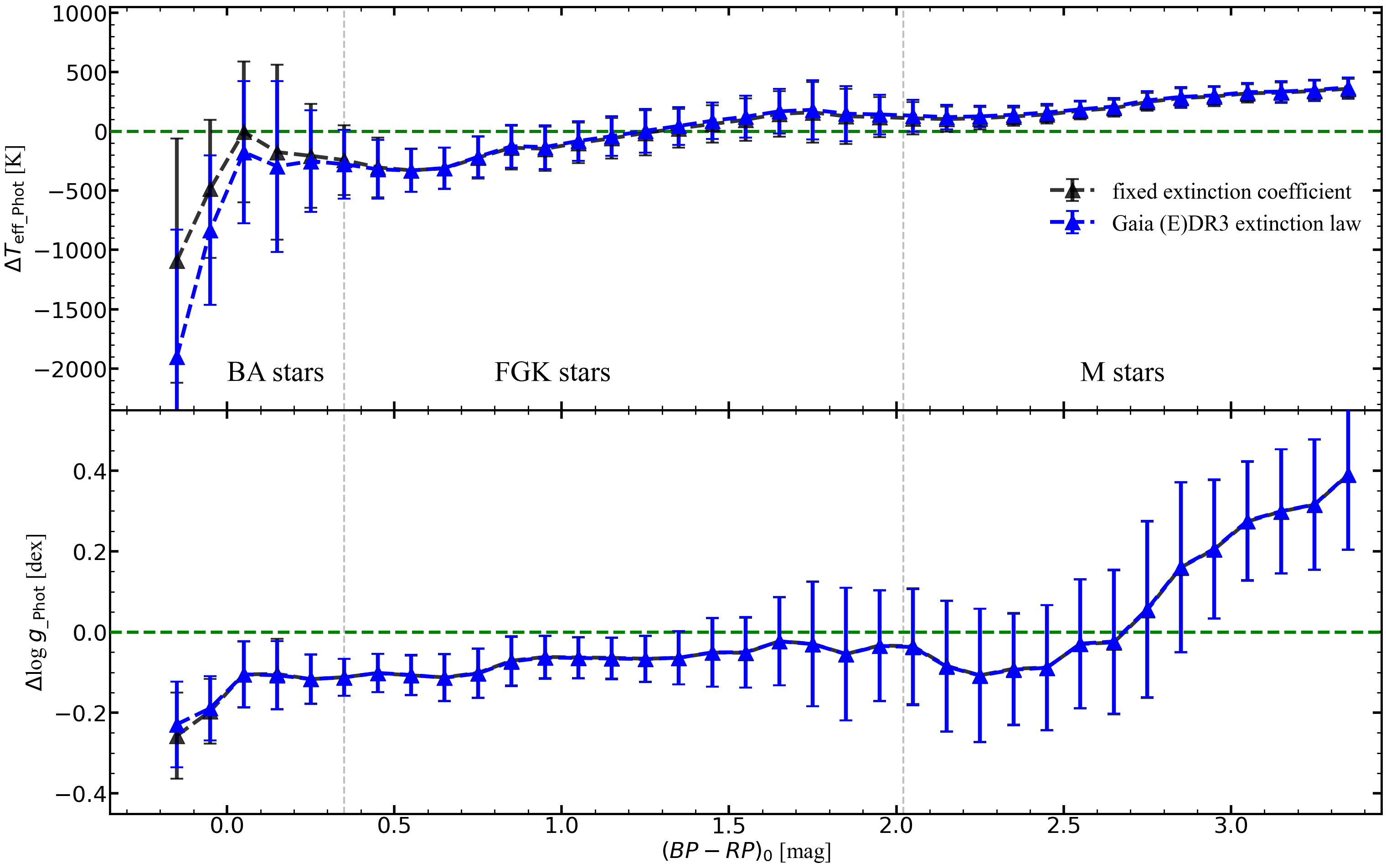}
\caption{Deviation distribution of the $T_{\rm eff}$ (upper panel) and log $g$ (lower panel) of GSP-Phot from the theoretical values obtained using different extinction laws. The dashed black and blue lines represent, respectively, the results obtained using the fixed extinction coefficient and {\it Gaia} (E)DR3 extinction law.}
\label{fig:D5}
\end{figure}

Consequently we applied the {\it Gaia} (E)DR3 extinction law to the member stars and recalculated the theoretical values of their atmospheric parameters, and the resulted distributions of $\Delta T_{\rm eff\_Phot}$ and $\Delta \log g_{\_Phot}$ are shown as the dashed blue lines in Fig.~\ref{fig:D5}. Compared with using a fixed extinction coefficient (dashed black line in Fig.~\ref{fig:D5}), for B and A-type stars, the $\Delta T_{\rm eff\_Phot}$ using the {\it Gaia} (E)DR3 extinction law is more severely underestimated. This is in fact because the theoretical $T_{\rm eff\_iso}$ estimated using the {\it Gaia} (E)DR3 extinction law is higher, which further causes the deviations to become larger. As for the $\Delta \log g_{\_Phot}$ deviation of B-type stars, there is an improvement to some extent, but this is not significant. In addition, for F, G, K, and M-type stars, the impact is rather negligible. This further indicates that the source of the deviation between the observed values of B and A-type stars and the theoretical values could be related to the method used.

\subsection{The results of GSP-Spec with different quality flags}\label{sec:D3}

In Sect.\ref{sec:gspspec}, we selected the samples with the best quality in GSP-Spec by applying the condition that the quality flags of vbroadT, vbroadG, vbroadM, vradT, vradG, and vradM are all equal to 0. That is, f1, f2, f3, f4, f5, and f6=0 in the corresponding gspspec\_flags. In this section, we discuss the deviation of the $T_{\rm eff\_Spec}$ and log $g_{\rm\_Spec}$ of the low-quality samples.

\begin{figure}[!htbp]
\includegraphics[width=0.48\textwidth]{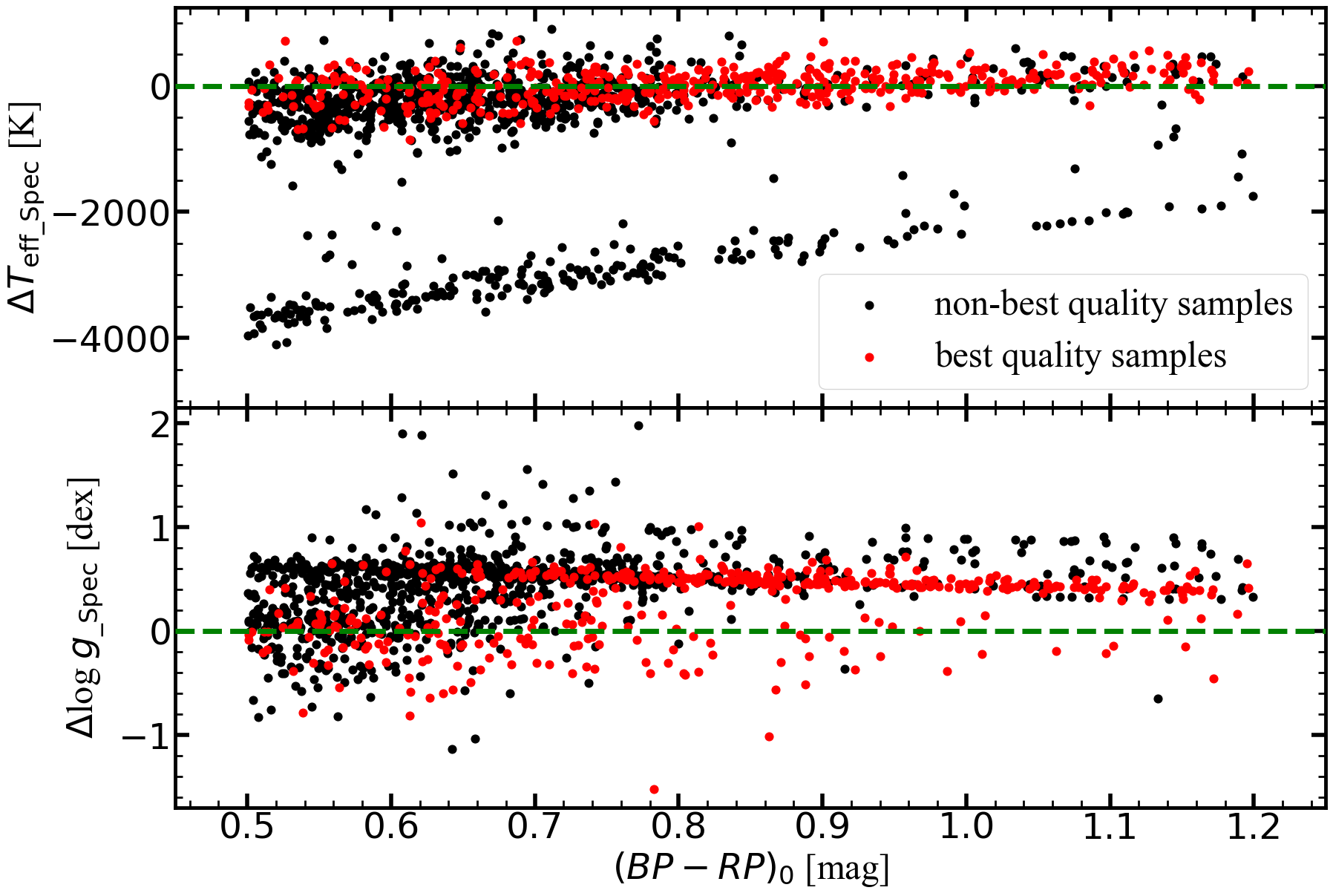}
\caption{Distribution of the deviations of $T_{\rm eff\_Spec}$ and log $g_{\rm\_Spec}$ for different quality flags in GSP-Spec. The red dots represent the highest-quality samples where f1, f2, f3, f4, f5, and f6 in the gspspec\_flags are all equal to 0, and the black dots represent the low-quality samples where at least one of the flags f1, f2, f3, f4, f5, and f6 is not equal to 0.}

\label{fig:D3}
\end{figure}

The black dots in Fig.~\ref{fig:D3} represent the deviation of $T_{\rm eff\_Spec}$ and log $g_{\rm\_Spec}$ of the member stars in GSP-Spec with low-quality flags, meaning that at least one of the quality flags f1, f2, f3, f4, f5, and f6 is not equal to 0. In contrast, the red dots denote the highest-quality samples. We have found that in the upper panel of Fig.~\ref{fig:D3}, there is a significant underestimation of the $T_{\rm eff\_Spec}$ for the member stars with low-quality flags. For a portion of the low-quality sample, the deviation of $T_{\rm eff\_Spec}$ increases linearly from $2000\,\rm K$ to $4000\,\rm K$. After the quality control, as is shown by red dots, this obvious linear structure of large deviation can be effectively filtered out. However, in the lower panel of Fig.~\ref{fig:D3}, we found that the deviation of the $\log g_{\_Spec}$ of the highest-quality samples is nearly the same as that of the low-quality samples; both exhibit overestimating tendencies.

\subsection{The limitations of our method}\label{sec:D4}

It is worth noting that there are still certain limitations in our work. We have used the theoretical parameters obtained from the cluster isochrones to validate the observed atmospheric parameters; however, we have not considered the fitting error of the cluster age, which may cause a certain degree of deviation in the theoretical parameters of the member stars. Although we removed binaries with a mass ratio greater than 0.5 using the binary sequence for each cluster, the low mass-ratio binaries still affect the theoretical parameter estimation. Moreover, for some clusters, the small color deviations between the observational CMDs and theoretical PARSEC isochrones exist, especially for the low-mass region \citep{2024ApJ...971...71J,2024arXiv241112987W}, which would impact the atmospheric assessment. In addition, the extinction coefficient we used is more applicable to the temperature range of [5250, 7000]\,\rm K. For some higher-temperature stars, different extinction laws can have significant impacts on the deviation of the observed values.

The uncertainties in the PARSEC model are not considered in our work, which can be divided into two parts: physical uncertainties and computational uncertainties \citep{2013A&A...549A..50V,2016A&A...586A.119S}. The physical uncertainties include nuclear reactions, radiative and conductive opacity, and mixing processes. The computational uncertainties include spatial and temporal resolutions in models, and stellar structure equation solutions. It is difficult to quantify the model uncertainties in $T_{\rm eff}$ and $\log g$, which is beyond the main goal of this article. It is important to emphasize that the isochrone models used do not include stellar rotation, which can introduce systematic offsets in the interpretation of atmospheric parameters, particularly for early-type stars.

\section{Conclusions}\label{sec:C}

In this paper, we selected 130 open clusters with clear and well-defined main sequences within $500\,\rm pc$ of our solar neighborhood to assess the quality of atmospheric parameters from large spectroscopic surveys such as {\it Gaia,} LAMOST, APOGEE, and GALAH. Meanwhile, we applied photometric quality filters and binary fraction criteria to remove sources with bad photometric observation or binary stars. Then, we estimated the theoretical atmospheric parameters for each cluster member based on the best isochrone fit by \citetalias{2023ApJS..265...12Q}. Utilizing this unified reference library of atmospheric parameters, we evaluated the quality of atmospheric parameters from GSP-Phot, GSP-Spec, and ESP-HS of {\it Gaia} DR3, as well as LAMOST-LRS DR11, APOGEE DR17, and GALAH DR4.

Although the work of using open cluster member stars as a reference standard for atmospheric parameters and then validating the atmospheric parameters of survey data has been carried out, our sample selection is more stringent and statistically significant. By choosing a higher-quality sample of open cluster members, we are better able to assess the typical biases and dispersions of atmospheric parameters in different survey data. For example, in comparison with the work of validating atmospheric parameters by \citetalias{2023A&A...674A..28F}, it is noted that differential reddening can introduce significant biases when estimating the theoretical atmospheric parameters. Therefore, we selected clusters with clear and well-defined main sequences and removed large mass-ratio binaries (q>0.5).

Our major results include:

(1) For B and A-type stars, there is an underestimation of $T_{\rm eff}$ of these surveys, and the dispersions in B and A-type stars are all relatively large. All surveys show a significant underestimation of log $g$, except for the 0.01\mbox{\,\rm}dex of APOGEE DR17. There is a systematic offset of $-$0.2\mbox{\,\rm}dex between ESP-HS and isochrone values. We need to point out that our method does not take into account the factor of rotation, which may introduce systematic effects. Moreover, the difference in extinction law may also have an impact on the results.

(2) For F, G, and K-type stars, the $T_{\rm eff}$ of the individual surveys are relatively consistent with the isochrones, and the dispersions are all within 260\mbox{\,\rm}K. The log $g$ of each survey shows an underestimation, except for GSP-Spec, where there is an overestimation of 0.43\mbox{\,\rm}dex. GSP-Phot and APOGEE DR17 show the systematic offset of $-$0.08\mbox{\,\rm}dex and $-$0.05\mbox{\,\rm}dex compared to the isochrone values.

(3) For M-type stars, the $T_{\rm eff}$ of GSP-Phot, APOGEE DR17, and GALAH DR4 all show a relatively large positive deviation, but the dispersions are smaller, with a dispersion of 61\,\rm K being the smallest for APOGEE DR17. The $\log g$ of APOGEE DR17 has a systematic offset of $-$0.03\mbox{\,\rm}dex compared to the isochrone values, and the dispersion is also minimized at 0.09\,\rm dex.

We expect this work to serve as a valuable reference for producers and users of large-scale surveys -- spectroscopic or photometric, from ground or space.

\section{Data availability}

Table~\ref{tab:appendix} is only available in electronic form at the CDS via anonymous ftp to cdsarc.u-strasbg.fr (130.79.128.5) or via \url{http://cdsweb.u-strasbg.fr/cgibin/qcat?J/A+A/}.

\begin{acknowledgements}
We express our gratitude to the anonymous referee for their valuable comments and suggestions, which are very helpful in improving our manuscript.
This work is supported by the National Natural Science Foundation of China (NSFC) through grants 12090040, 12090042. 
Jing Zhong would like to acknowledge the science research grants from the China Manned Space Project with NO. CMS-CSST-2025-A19, the Youth Innovation Promotion Association CAS, the Science and Technology Commission of Shanghai Municipality (Grant No.22dz1202400), and Sponsored by the Program of Shanghai Academic/Technology Research Leader.
Li Chen acknowledges the science research grants from the China Manned Space Project with NO. CMS-CSST-2021-A08.
Songmei Qin acknowledges the financial support provided by the China Scholarship Council program (Grant No. 202304910547).
This work has made use of data from the European Space Agency (ESA) mission {\it Gaia} (\url{https://www.cosmos.esa.int/gaia}), processed by the {\it Gaia} Data Processing and Analysis Consortium (DPAC, \url{https://www.cosmos.esa.int/web/gaia/dpac/consortium}). Funding for the DPAC has been provided by national institutions, in particular, the institutions participating in the {\it Gaia} Multilateral Agreement.
\end{acknowledgements}

\bibliographystyle{aa}
\bibliography{ref.bib}

\begin{appendix}
\section{Reference atmospheric parameters catalog}{\label{Appendix}}

We provide the catalog with atmospheric parameters of the member stars used for reference. A complete list of these stars is available at the CDS, and the description of the catalog is shown in Table.~\ref{tab:appendix}. Columns $(1)-(10)$ list fundamental information about the member stars, including the Name of the cluster to which it belongs (Name), {\it Gaia} DR3 source identifier (gaia\_source\_id), the position coordinate in ICRS (RAdeg, DEdeg), and the $G$, $BP$, and $RP$ magnitude and their errors in {\it Gaia} DR3. Columns $(11)-(14)$ list the atmospheric parameters ($T_{\rm eff\_iso}$, log $g_{\rm\_iso}$) we estimated by matching the isochrones, and the corresponding uncertainty (e\_$T_{\rm eff\_iso}$, e\_log $g_{\rm\_iso}$). Column (15) list the minimum difference between the observed and theoretical values of $\sqrt{\Delta(BP-RP)_0^2+\Delta M_{\rm G}^2}$. A positive value indicates that the member star is on the right side of the isochrone, while a negative value indicates that it is on the left side.

We sampled 50 repetitions of the normality of $Gmag\sim\mathcal N(Gmag, Gmag\_{err}^{2})$ and $(BP-RP)\sim\mathcal N((BP-RP),\sqrt {BPmag\_{err}^{2}+RPmag\_err^2})$ for each member star, and used the standard deviation of the atmospheric parameters obtained from these samples as the uncertainty (e\_$T_{\rm eff\_iso}$, e\_log $g_{\rm\_iso}$). 
It is important to note that our estimates of uncertainties are underestimated. This is because our estimation is only based on the observations' uncertainties, and we do not consider the uncertainties due to age, and stellar evolution models.

\begin{table}[!htbp]
\renewcommand{\arraystretch}{1.3}
\caption{Description of the catalog of the member stars we used.}\label{tab:appendix}
\centering
\Large 
\scalebox{0.7}
{
\begin{tabular}{lcc}
\hline\hline
Column                  & Unit         & Explanations\\
\hline
Name                    &  -           & Cluster name from \citetalias{2023ApJS..265...12Q}  \\
gaia\_source\_id        &  -           & Unique source identifier, {\it Gaia} DR3   \\
RAdeg                   &  deg         & Right ascension, {\it Gaia} DR3  \\
DEdeg                   &  deg         & Declination, {\it Gaia} DR3  \\
Gmag                    &  mag         & G magnitude, {\it Gaia} DR3   \\
Gmag\_err                 &  mag         & G magnitude error, {\it Gaia} DR3   \\
BPmag                   &  mag         & BP magnitude, {\it Gaia} DR3   \\
BPmag\_err                &  mag         & BP magnitude error, {\it Gaia} DR3   \\
RPmag                   &  mag         & RP magnitude, {\it Gaia} DR3   \\
RPmag\_err                &  mag         & RP magnitude error, {\it Gaia} DR3   \\
$T_{\rm eff\_iso}$      &  K           & Estimated $T_{\rm eff}$  from isochrones   \\
e\_$T_{\rm eff\_iso}$ &  K           & Uncertainty in $T_{\rm eff\_iso}$   \\
log $g_{\rm\_iso}$       &  dex         & Estimated log $g$  from isochrones   \\
e\_log $g_{\rm iso}$  &  dex         & Uncertainty in log $g_{\rm iso}$  \\
min\_distance  &  mag         & The minimum distance  \\

\hline
\end{tabular}
}
\end{table}
\end{appendix}
\end{CJK*}

\end{document}